\documentclass[11pt,twoside]{article}
\usepackage{newpasp}
\usepackage{epsf}

\index{Braun, R.}

\def\Msun{\hbox{$M_\odot$}}
\def\HI{\hbox{H {\sc i}}}
\def\HII{\hbox{H {\sc ii}}}
\def\CO1_2{\hbox{CO$1\rightarrow 2$}}
\def\H2{\hbox{H$_2$}}
\def\kms{km\,s$^{-1}$} 

\begin{document}

\title{Compact High Velocity Clouds}
\author{Robert Braun}
\affil{Netherlands Foundation for Research in Astronomy, P.O.Box 2,
  7990 AA Dwingeloo, The Netherlands} 


\begin{abstract}
  
  We summarize the observed properties of the CHVC population, which
  provide strong evidence for source distances in the range
  200--1000~kpc. At these distances, the population 
  corresponds to strongly dark-matter dominated sub-dwarf galaxies
  still accreting onto the more massive Local Group systems.  Recent
  searches for faint associated stellar populations have revealed
  red-giant candidates for which follow-up spectroscopy is scheduled. A
  sensitive \HI\ survey for CHVC counterparts in the NGC~628 galaxy
  group has allowed tentative detection of 40 candidates, for which
  confirming observations have been approved.  Many open issues should
  be resolved by observational programs within the coming years.

\end{abstract}




\section{Introduction}

The \HI\  High Velocity Cloud (HVC) phenomenon has been studied for some
three decades, but it is only with the sensitive, all-sky imaging of
the LDS (Hartman \& Burton 1997) in the North and HIPASS
(Staveley-Smith et al 1999) in the South, that some important aspects
of the phenomenon are becoming apparent. Even a casual inspection of
these data reveal a striking dichotomy in the types of emission
features present. The vast majority of the emission flux is associated
with diffuse filamentary complexes extending over 10's of degrees, of
which the Magellanic Stream (Putman 2000) is a prime example. A much
smaller total flux is associated with a population of compact,
high-contrast features which are isolated from any other \HI\  emision
features both spatially and in velocity (the CHVCs). My attention was
first drawn to this second object class, while using the LDS data to
assess possible Galactic foreground confusion in the directions of
nearby galaxy groups in the fall of 1997. I was surprised to see high
contrast peaks of uncataloged high-velocity \HI\  which were
indistinguishable in the LDS data from the \HI\  signatures of nearby
dwarf galaxies. This was so intriguing to me that I set about making a
catalog of such objects based on the LDS data (covering
$\delta~=~-30$--90$^\circ$).  Together with Butler Burton, follow-up
observations of all candidate sources were obtained and a list of 65
confirmed objects were published (Braun \& Burton 1999).

\section{CHVC Properties}

In addition to having a distinctive, compact morphology (average FWHM
of 50$\pm$25 arcmin) the CHVCs also proved to have a very systematic
spatial and kinematic distribution on the sky. A global search over all
direction cosines for the reference system that minimized the velocity
dispersion of the population, returned the Local Group Standard of Rest
(LGSR) with high significance. In the LGSR, the CHVC population has a
velocity dispersion of only 70 \kms, in contrast to 95 \kms\  measured in
the Galactic Standard of Rest (GSR). But rather than being at rest in
the LGSR frame, the entire system is infalling by about 100 \kms\  toward
the Local Group barycenter. These distinctive kinematic properties of
the CHVC population were the first indication that they might reside at
substantial distances, and were not merely a local ISM or Galactic
phenomenon.

\subsection{Velocity Gradients and Multi-Core Systems}

Further indications for a substantial distance to these objects have
emerged from a program of high resolution \HI\ synthesis imaging with
the WSRT (Braun \& Burton 2000) and in total power with Arecibo
(Burton, Braun \& Chengalur 2000). The general picture which has
emerged is that of one or more compact cores (of 1--20~arcmin extent)
embedded in a common diffuse halo. On average about 40\% of the \HI\ 
line flux is due to the compact cores, although this varies enormously;
ranging from $<$1\% for the lowest flux objects (less than about
50~Jy-\kms) to about 50\% for brighter systems (100--300~Jy-\kms). Many
of the compact cores display velocity gradients (of 10--30 \kms\ 
amplitude) along the long dimension of an approximately elliptical
extent. For multi-core objects, each core has it's own distinct
kinematic axis, which is uncorrelated with that of other cores in the
same object (and can be perpendicular as often as parallel). It is
difficult to envision an external mechanism (such as tidal shear for
example) for imparting vastly different kinematic axes to the various
cores of a single object.  Resolved detection of the diffuse halos with
deep Arecibo imaging has shown that the halos do not participate in the
same velocity gradients of the cores, but are more nearly stationary at
the systemic velocity of the object and always have the 22+~\kms\ FWHM
linewidth of an 8000+~K gas.  This implies that either (1) the velocity
gradients of the cores have an internal (like self-gravity) rather than
external origin or (2) the objects are sufficiently long-lived that the
halos can become thermalized (and so erase their original kinematics).
In any case, the thermalization timescale for a 50~arcmin diffuse halo
is $\tau_t \sim D_{kpc}$~Myr, at a distance of $D_{kpc}$ in kpc.

The most extreme multi-core CHVC yet studied (CHVC115+13$-$275) has
some 10 major core components spanning about 30~arcmin on the sky with
centroid velocities distributed over no less than 70~\kms!  This object,
and others like it, either have an extremely short dynamical lifetime
($\tau_d < 0.1D_{kpc}$~Myr) or are bound by a large amount of dark matter.
From the very narrow \HI\ linewidths seen in the cores we know that
internal conditions are appropriate for maintaining \HI\ at about
100~K, which requires that the product, $n_H t > 1$ cm$^{-3}$Myr
(Draine 1978).

While self-gravity could account for all of these facts in a natural
way: (1) the cool \HI\ cores by the enhanced internal pressure of a
self-gravitating system, (2) the local velocity gradients as rotation
in dark-matter-dominated mini-halos, (3) the binding of multi-core objects
by a massive dark halo; it is not clear which alternative
scenario might work. The projected velocity fields of the best-resolved
cores can be well-modeled by a slowly rising rotation curve, leveling
off at 8~arcmin radius to a velocity of 15--20~\kms. The implied total
masses in these individual cores is about $D_{700}10^8$\Msun, while for
the high linewidth multi-core systems it is nearer $D_{700}10^9$\Msun,
in terms of an assumed distance $D_{700}$ normalized to 700~kpc. Even
at these rather substantial assumed distances, the dark-to-visible mass
ratios are quite high with individual cores ranging from,
$\Gamma$ = 10--40/$D_{700}$, while for the multi-core systems $\Gamma
\sim$ 50/$D_{700}$.

\subsection{Opaque \HI\ Clumps}

The objects with the narrowest \HI\ emission lines have also proven
particularly interesting. Compact clumps of opaque \HI\ (with 90~arcsec
FWHM) have been observed within the cores of at least one source to
date (CHVC125+41$-$207). Such clumps have \HI\ brightness temperatures
of 75~K, together with FWHM linewidths of less than about 2~\kms. These
values allow a reliable determination of the kinetic temperature (85~K)
and opacity ($\tau\sim$2) as well as an upper limit on turbulent
contributions to the internal linewidth ($<$1~\kms). Since we have
measured an angular size and excess column density for the clumps we can
estimate the distance from $D~=~dN_H/(n_H\theta)$, assuming crude
spherical symmetry, if we have an estimate of the \HI\ volume density,
$n_H$. 

We can perhaps demonstrate this method best by using M31 as an example. The
excess column density, $dN_H = 5\pm1\times10^{21}$cm$^{-2}$, and
angular size, $\theta = 60\pm20$~arcsec, of \HI\ clumps in the
North-East half of M31 can be estimated from Fig.~5 of Braun \&
Walterbos (1992). The \HI\ kinetic temperature, $T_k = 175\pm25$~K, and
estimated thermal pressure, $P/k = 1500\pm500$~cm$^{-3}$K, of the M31 mid-disk
are tabulated in Table~4 of the same reference. The implied volume
density, $n_H = 9\pm3$~cm$^{-3}$, leads to the distance estimate:
$D_{M31} = 670\pm220$~kpc. While rather crude, this approach gives a
plausible value for the distance to M31. For CHVC125+41$-$207, we have an
excess column density, $dN_H = 1.0\pm0.2\times10^{21}$cm$^{-2}$, and
angular size, $\theta = 90\pm15$~arcsec, of the \HI\ clumps for which
$T_k = 85\pm10$~K.  Although the thermal pressure, $P/k$, in the solar
neighborhood is about 2000~cm$^{-3}$K, it is expected to decline
rapidly with distance from the Galactic plane (eg. Wolfire et al. 1995)
such that beyond about 10~kpc we should encounter values, $P/k =
150\pm50$~cm$^{-3}$K, leading to an estimated volume density, $n_H =
2\pm1$~cm$^{-3}$. In fact, as described in Braun \& Burton (2000), an
\HI\ temperature equilibrium calculation was performed by Wolfire et
al. (2000) assuming a Local Group radiation field and a metallicity and
dust-to-gas ratio of 10\% Solar which allowed exactly these pressure
and density estimates to be made for this object. The resulting
distance estimate is, $D_{C125} = 600\pm300$~kpc.

\begin{figure}[t!]
\plottwo{braunr_fig1.tex}{braunr_fig2.tex}
\vspace*{-0.4cm}
\caption{Edge profiles of the indicated CHVCs. The logartithm of
  \HI\ column density is plotted against distance from the source
  centroid for both the Eastern and Western halves of the source. The
  dotted curve is the sky-plane projection of a spherical exponential
  with the parameters shown. 
}
\vspace*{-0.4cm}
\end{figure}

\subsection{Halo Edge Profiles}

Another result of the Arecibo imaging of CHVCs reported in Burton,
Braun \& Chengalur (2000) was the detection of approximately
exponential edge profiles of the diffuse halos between column densities
of about 10$^{19}$ down to the limiting sensitivity (1$\sigma$ over
70~\kms) of $2\times10^{17}$cm$^{-2}$. (BTW, these are probably some of
the deepest \HI\ emission measurements of resolved objects yet
obtained.)  The edge profiles are well-fit by the sky-plane
projection of a 3-D spherical exponential in \HI\ volume density,
$n_H(r) = n_o e^{-r/h}$, in terms of the radial distance, $r$, and
exponential scale length, $h$, yielding $ N_H(r) = 2 h n_o (r/ h) K_1 (
r/h) $, where $K_1$ is the modified Bessel function of order 1. Fitting
this form to the observed edge profiles allows accurate assessment of
both the central column density of the warm halo, $N_H(0)$, and the
scale length, $h$, in the plane of the sky. As before, with any
reasonable estimate of the thermal pressure, it becomes possible to
determine the source distance assuming crude spherical symmetry.
Assuming a pressure in the diffuse CHVC halos, $P/k \sim
100$~cm$^{-3}$K, results in distance estimates which vary between
150--1100~kpc for the ten CHVCs with measured edge profiles.

\subsection{Do LMC Distances Work?}

What about placing the CHVCs at the distance of the LMC/SMC? All
existing estimates of the thermal pressure at these distances lie at or
below about $P_{th}/k\sim 100$~cm$^{-3}$K. To this could be added the
ram pressure $P_{ram} \sim \rho v^2/2$ of passage through the hot halo
of the Galaxy. Assuming a typical relative velocity of 100~\kms, and a
Galactic halo temperature of 10$^6$K, gives $P_{ram} = 0.8 P_{th}$,
implying a significant, but not dominant contribution to the total
pressure.  In order to account for the \HI\ column densities and
angular sizes of both the opaque clumps seen in CHVC125+41$-$207 as
well as the diffuse halos of the ten CHVCs observed with Arecibo, we
would require some additional source of pressure to yield 3--20 times
higher values. It is not clear where this might come from. At a
distance of about 50~kpc, the thermalization timescale of the diffuse
CHVC halos is 50~Myr, while the dynamical lifetime of
multi-core CHVCs like CHVC115+13$-$275 is only 5 Myr. So, on the one
hand, the absence of velocity gradients in the diffuse halos demands a
long source lifetime, while on the other, the sources are too
short-lived for the halos to be thermalized. This scenario is very
difficult to reconcile with the observed CHVC properties.

\section{Current Work on CHVCs}

Although a large number of the CHVC attributes suggest Local Group
distances, the various distance determinations outlined above remain
indirect. Essentially, a self-consistent scenario has emerged, in which
many physical and kinematic properties of the CHVCs can be understood
if they are self-gravitating, dark-matter-dominated systems.  It is
very appealing to associate them with the low-mass ``building-blocks''
of galaxy formation (Blitz et al. 1999), which high resolution
numerical simulations suggest should still be found in large numbers in
the appropriate environments (Klypin et al. 1999, Moore et al.  1999).
If this connection can be demonstrated convincingly we would have the
opportunity to gain important insights into a wide range of fundamental
problems in astrophysics, through the study of these cosmological
fossils in our own ``backyard''.  However, demonstrating consistency
does not constitute a proof of the conjecture. We are pursuing several
different lines of research to clarify the nature of these objects.

\subsection{All-Sky CHVC Population}

Vincent de Heij (as part of his PhD work in Leiden) working together
with Mary Putman, has just completed cataloging the population of
Southern hemisphere CHVCs using the HIPASS data (Putman, De Heij et al.
2000). The Southern hemisphere data is particularly important in
constraining the kinematics of the population, since this is where it
extends to the highest positive velocities in the LSR reference frame.
It is also special in the sense that both the nearest external
galaxies, the Magellanic Clouds, as well as the nearest external galaxy
group, the Sculptor Group, are located there. Just as was seen
previously in the LDS data North of $\delta=-30^\circ$, there is a
clear distinction in the HIPASS data between the extended filamentary
complexes which make up the bulk of peculiar velocity \HI\ emission and
the compact, isolated CHVCs, which have the \HI\ appearance of nearby
dwarf galaxies. About 100 well-defined CHVCs were cataloged below
$\delta=0^\circ$ in the HIPASS data. A new analysis of the spatial and
kinematic properties of the all-sky CHVC population will be carried out
in the coming months. High resolution \HI\ imaging data has also been
acquired with the WSRT for eight additional objects. These targets were
chosen on the basis of high brightness temperatures in the LDS data, in
the hope of detecting more opaque clumps of the type seen in
CHVC125+41$-$207.

\subsection{Stars in CHVCs}

An unambiguous distance determination for the CHVCs would follow from
direct detection of an associated stellar population. A high surface
brightness stellar disk can already be ruled out for all of the CHVCs
outside of the Galactic plane Zone of Avoidance. The constraints on low
surface brightness populations are not yet very strong, especially
since they would be highly resolved into individual stars at distances
of 200--1000~kpc, which would be difficult to distinguish from the
dense Galactic foreground. If young, high mass stars were present, then
these should give rise to prominent \HII\ regions. Together with Ren\'e
Walterbos (NMSU), we have begun a program of narrow-band H$\alpha$
imaging of CHVCs with the APO 1-m; so far without detections in the
first three fields imaged. An older stellar population could be traced
by its most luminous and populous component, namely the red-giant
branch. Deep searches for associated red-gaints have been carried out
in the spring and summer of 2000 using the Mosaic cameras on the KPNO
4-m and CTIO 4-m together with Eva Grebel and Daniel Harbeck (MPIA) and
the LCO 2.5-m together with Carme Gallart (Yale/Chile) and Steve
Majewski (UVa). Besides several broad-band colors, the DDO51 filter,
centered on the MgH absorption feature at 5100 \AA, was used to allow
reliable discrimination of the many foreground white-dwarfs from faint
background objects. While still only partially analyzed, these data do
suggest a sparse population of candidate stars in several objects with
the colors of red-giants and luminosities consistent with a 700~kpc
distance. However, at these faint brightnesses, there is the
possibility of confusion by background galaxies at red-shifts of a few
tenths which resemble the sought for red-giant population when viewed
with typical ground-based angular resolutions of about 1~arcsec. Two
additional observing programs together with Eva Grebel and Kem Cook
(LLNL) have already been allocated for fall 2000 to clarify this issue;
namely high resolution imaging with the VLT, and multi-object
spectroscopy of candidate stars with Keck. Stay tuned \dots

\subsection{Exo-CHVCs}

Another way to circumvent the difficulties of distance determination of
the Local Group CHVCs is to detect CHVC counterparts in external groups
of galaxies. However, the low mass end (below M$_{HI}~\sim~10^8$
M$_\odot$) of the \HI\ mass function (HIMF) is still a topic of very
active current research. Only small numbers of low mass objects have
been detected to date in blind, large area surveys: eg. 4 out of 66 in
the AHISS (Zwaan et al. 1997), 4 out of 79 in the ``Arecibo
Slice'' (Spitzak \& Schneider 1998, ) 3 out of 263 in the HIPASS
(Kilborn et al. 1999) and 7 out of 265 in the ADBS (Rosenberg \&
Schneider 2000).  Consequently, there are substantial uncertainties in
the space densities at low HI mass due to small number statistics.

Targeted searches for uncataloged objects in specific environments are
beginning to supplement the blind surveys. Banks et al. (1999) detect
17 out of 27 objects with M$_{HI}~<~10^8$ M$_\odot$ in the Cen A group
of galaxies using HIPASS data. Verheijen et al. (2000)
detect comparable numbers of objects in the Ursa Major cluster in a
recent VLA survey.

Comparison of the HIMFs derived by the various authors is beginning to
show what might be substantial differences in the statistics of the low
mass populations seen in different environments. Within the Ursa Major
cluster, for example, the low mass HIMF is flat or even declining
towards lower masses. In the poorer, but still relatively rich,
environment of the Cen A group there appears to be a significant
increase to lower masses. The detected space density in the lowest
measured bin (11 objects centered at $2\times10^7$ M$_\odot$) is twice
that seen in the next lowest bin (6 objects centered at $6\times10^7$
M$_\odot$). A similar, but still low significance, upturn is apparent
in the population of ``field'' galaxies sampled by both the Southern
sky HIPASS data and the combined AHISS and Arecibo Slice data reported
by Schneider et al. (1998).

Detection of rather different HIMFs in different environments is not in
itself a surprising result. The interaction, merger and stripping rates
of low mass systems must depend sensitively on the richness of their
environment. Optical luminosity functions have not demonstrated a
strong dependence of shape with environment (eg. de Propris et al. 1995;
Loveday 1997; Marzke et al. 1998) but what they all seem to share is a
strong upturn in the space density of low luminosity systems below $M_B
\sim -$15. Given the much larger interaction cross-section of diffuse
gas over that of stars, we should expect dramatic differences in the
low mass end of the HIMF, even if the ``primordial'' HIMF were the same
everywhere.

The latest HI surveys have begun to reveal a rather interesting
population of sources at the very low mass end in poor environments.
The previously uncataloged \HI\ detections in the Cen~A group (Banks et
al. 1999) all have very low luminosity ($M_B \sim -$11), low surface
brightness $<\mu> \sim$ 26 mag arcsec$^{-2}$ optical counterparts. In
the case of the ADBS (Rosenberg \& Schneider 2000) fully 22 of 81
previously uncataloged objects have no clear optical counterpart down
to the limiting magnitude of the POSS (about 25 mag arcsec$^{-2}$). Of
these 22 objects, 11 have substantial extinction (A$_V >$ 2 mag) and
therefore do not provide strong limits, but 5 have negligible
extinction (A$_V <$ 0.3 mag) and are thus quite interesting. The most
extreme object in this category may be the possible Cen~A group member
HIPASS J1712$-$64 (Kilborn et al. 2000) for which an
upper limit of $\mu_B$ = 27 mag arcsec$^{-2}$ is associated with an HI
mass of $1.7\times10^7$ M$_\odot$.
Are we perhaps encountering the tip of an iceberg of very low mass
systems with only a faint (or even no) population of associated stars?

Together with Butler Burton, we have undertaken an attempt to probe the
extreme low mass end of the HIMF in a poor environment by carrying out
a deep survey of the NGC~628 galaxy group. Nineteen sparse pointings,
distributed over a region of about 1~Mpc (or 5 degrees) in diameter,
were each observed with a 12 hour integration using the recently
upgraded Westerbork array. The 5$\sigma$ HI mass sensitivity at each
field center was $2\times10^6$ M$_\odot$ over the minimum expected
linewidth of 32~km~s$^{-1}$ FWHM (although the velocity resolution was
5~km~s$^{-1}$). Since the pointings were sparsely distributed, they
only overlap at the 5\% level of the circular Gaussian primary beam
with 2100 arcsec FWHM. At these ``cracks'' in the hexagonal pointing
pattern, the 5$\sigma$ mass sensitivity is degraded to $4\times10^7$
M$_\odot$ over 32~km~s$^{-1}$.  Applying a significance criterion of
0.5 random noise detections per data cube (varying between 5.5 and 5.0
times $\sigma$ depending on the degree of velocity smoothing, and hence
the number of independent pixels) resulted in a list of 48 tentative
detections in the complete survey. Of these, only 7 correspond to
previously cataloged galaxies. One of the remaining 41 candidates has
an obvious optical counterpart in the POSS data, while three more have
faint wisps in the POSS that may be low surface brightness optical
counterparts (with peak $\mu_R \sim$ 25 mag arcsec$^{-2}$). The
remaining 37 candidates have no optical counterparts down to this
surface brightness limit. Given the noise statistics we expect about
10 of our candidates to be random noise peaks, which is also the number
of ``negative sources'' found with the same criteria in the survey data.
Confirming observations of our low mass candidates with the Arecibo
telescope have been approved for fall 2000. 

\section{Conclusions}

These are very interesting times for high velocity \HI\ cloud
research. Substantial progress has been made in determining the
physical properties of these objects. The weight of current evidence
seems to favor a large mean distance for the compact isolated objects,
the CHVCs. More insights and perhaps definitive evidence should follow
within the next year or two from  a number of ongoing programs.




\acknowledgements

It is a pleasure to acknowledge the valuable input and collaboration of
many individuals in the work described here: Butler Burton, Jayaram
Chengalur, Vincent de Heij and Mary Putman for \HI\ aspects and Kem
Cook, Carme Gallart, Eva Grebel, Daniel Harbeck, Steve Majewski and
Ren\'e Walterbos for the optical program. This work could not have been
done without the generous allotment of telescope time by the ESO, Keck,
LCO, NAIC, NOAO, and NFRA time allocation committees.

\end{document}